\documentclass[prb,preprint]{revtex4}

\pdfoutput=1

\usepackage{graphicx}
\usepackage{latexsym}
\usepackage{amssymb}

\begin{document}


\title{Atomic configuration in cuprates at the closing of the pseudogap}

\medskip 

\date{October 5, 2021} \bigskip

\author{Manfred Bucher \\}
\affiliation{\text{\textnormal{Physics Department, California State University,}} \textnormal{Fresno,}
\textnormal{Fresno, California 93740-8031} \\}

\begin{abstract}
Doped holes in cuprates reside pairwise at lattice-defect $O$ atoms but at different sites in the two cuprate families. In the $Sr$-doped lanthanum cuprates, the $O$ atoms occupy anion lattice sites and spread due to Coulomb repulsion (relative to the host lattice). In the oxygenated cuprates, the $O$ atoms occupy interstitial sites, hybridize to ozone complexes, and aggregate to strings or (at high oxygenation) to nematic patches, always with spacing of $\sim 3.25a_0$. At the closing of the pseudogap at hole doping $p^*$, a highly symmetric configuration of the $O$ atoms appears in each family. In the first family it consists of interlaced superlattices in the $CuO_2$ plane and each of the bracketing $LaO$ layers, with spacing $A_0^{LaO}(p^*) = 2A_0^{CuO_2}(p^*)$. In the second family it consists in the completion of a 2D superlattice of ozone complexes, with spacing $A_0(p^*) \simeq 3.25a_0$. Both cases are visualized. Implications for the bandstructure and strange-metal phase are considered.

\end{abstract}

\maketitle

\section{TWO FAMILIES OF SUPERCONDUCTING CUPRATES}

It is useful to distinguish \textit{two} families of superconducting copper oxides (cuprates): those doped with heterovalent metals and those that are \textit{oxygenated}, that is, doped or enriched with oxygen. In the first family, ions of trivalent lanthanides, $Ln = La, Pr, Nd, Eu, ...$, of the host crystal are partially replaced by ions of divalent alkaline-earth, $Ae = Sr, Ba$, or of tetravalent cerium, $Ce$. This gives rise to a concentration of holes, $p = x$, and, respectively, electrons, $n=x$, in the $CuO_2$ planes. Examples are $La_{2-x}Sr_xCuO_4$ and $Nd_{2-x}Ce_xCuO_4$. The cuprates of the second family are either stoichiometrically doped with oxygen, as in $YBa_2Cu_3O_{6+y}$, or (non-stoichiometrically) enriched with oxygen as in $Bi_2Sr_2CaCu_2O_{8+\delta}$, $Tl_2Ba_2Cu_2O_{6+\delta}$, $HgBa_2CuO_{4+\delta}$, $La_2CuO_{4+\delta}$, and similar compounds.
In the case of $YBa_2Cu_3O_{6+y}$, the stoichiometric fraction $y$ is achieved by corresponding proportions of the chemical ingredients before crystal growth. In the other cases, oxygen enrichment $\delta$ is achieved after crystal growth by thermal treatment in an atmosphere of excess oxygen. In all oxygenated cuprates the density of holes in the $CuO_2$ plane, $p$, is determined by measurement of the onset temperature of superconductivity, $T_c$, or of the lattice constant $c_0$ perpendicular to the $CuO_2$ planes. The compound $Bi_2Sr_{1-x}La_xCuO_{6+\delta}$ is both oxygen enriched \textit{and} doped with heterovalent metal, resulting in the insertion of holes and (less) \textit{electrons} to the $CuO_2$ planes. 
Being more complicated, this material is not discussed here.

Both crystal families are commonly regarded as \textit{hole doped} (in the $CuO_2$ planes)---in the first family as a consequence of $Ae$-doping, in the second family as a consequence of oxygenation. However, there is an important difference between the two families that goes beyond the concentration of holes, resulting in very different crystal properties with regard to stripes and superconductivity. When in the compounds of the first family a doped alkaline-earth ion, $Ae \rightarrow Ae^{2+} + 2e^-$, replaces a host ion in the $LaO$ layer, $La \rightarrow La^{3+} + 3e^-$, the missing electron in the $LaO$ layer is taken from the $CuO_2$ plane, leaving a hole, $e^+$, behind. It can be shown that two such holes reside together at an anion lattice site of the $CuO_2$ plane, $2e^+ + \;O^{2-} \rightarrow O$.\cite{1} 
Relative to the host lattice, those $O$ atoms appear as charged entities. Coulomb repulsion spreads the lattice defects to a quadratic 2D superlattice of $O$ atoms with a doping-dependent spacing, $A_0(x)$, in each $CuO_2$ plane. Its reciprocal value is called the ``incommensurabiliy,'' $1/A_0(x) \equiv q_{CO}^{CuO_2}(x)$. The localization of the holes at lattice-site oxygen atoms, and their double occupancy, is inferred from the observed square-root dependence of the incommensurability of charge-order stripes on doping, $q_{CO}^{CuO_2}(x) \propto \sqrt{x}$, as well as of magnetization stripes, $q_M^{CuO_2}(x) = \frac{1}{2} q_{CO}^{CuO_2}(x)$.\cite{1}

In oxygen-doped $YBaCu_3O_{6+y}$ and the oxygen-enriched $Bi$, $Tl$ or $Hg$ cuprates, neutral oxygen atoms reside at interstitial sites in the $CuO_2$ planes, here denoted as $\mathring{O}$. Each one loosely bond with its neighboring $O^{2-}$ ions of the host lattice (here abbreviated as $O^{2-} \equiv \ddot{O}$) to form an ozone-ion complex, $O^{2-} + \mathring{O} + O^{2-} \rightarrow \ddot{O}\mathring{O}\ddot{O}$. 
Because of its interstitial site, the  $\mathring{O}$ atom appears \textit{neutral} relative to the host lattice. Accordingly, the ozone complexes don't spread but aggregate to strings, $...\ddot{O}\mathring{O}\ddot{O}$-$\ddot{O}$-$\ddot{O}\mathring{O}\ddot{O}$-$\ddot{O}$-$\ddot{O}\mathring{O}\ddot{O}...$, with a repetition length of 

\includegraphics[width=6.25in]{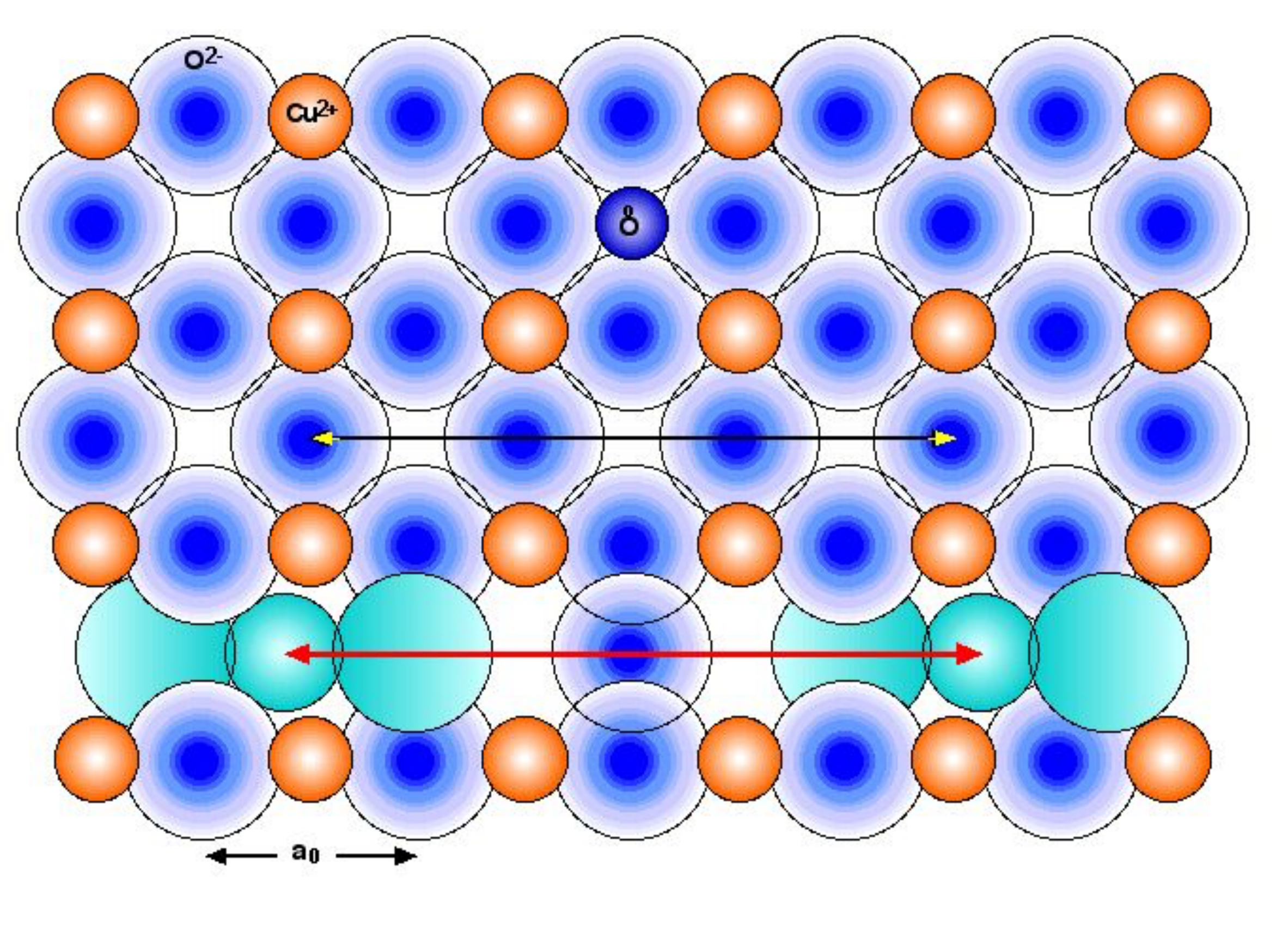}
\footnotesize 

\noindent FIG. 1. Oxygenated $CuO_2$ plane, shown schematically for oxygen atoms at the stage of embedding (second row) and of hybridization to ozone complexes (penultimate row). In the interstitial space between four $O^{2-}$ ions, the embedded oxygen atom (denoted $\mathring{O}$) bonds with two $O^{2-}$ neighbors, here in the $a$-direction, to form an ozone molecule ion (turquoise color).
The ozone complexes 
are slightly tilted out of the $CuO_2$ plane, as elaborated in Ref. 2.
They line up, with an intermediate $O^{2-}$ ion, to strings of periodic length $L \simeq 3.25\; a_0$ (red arrow), slightly larger than $3a_0$ (black arrow). Its reciprocal gives the incommensurabiliy of charge-order stripes, $q_{CO} =1/L \simeq 0.31$ r.l.u. \normalsize

\noindent about three and a quarter host-lattice constants,  $L \simeq 3.25 a_0$, (see Fig. 1). Its reciprocal value, $1/L = q_{CO}^{CuO_2} = 0.31 \pm 0.01$ r.l.u. (in reciprocal lattice units) is essentially a \textit{constant} incommensurability, independent of oxygen doping $y$ or enrichment $\delta$, as observed in  experiment. A weak dependence of the incommensurability on doping or axial direction in $YBa_2Cu_3O_{6+y}$, explained in Ref. 2, should not distract from the outstanding feature that $L \simeq 3.125 a_0$ seems to be the natural spacing of aggregated ozone complexes in oxygenated cuprates. As the ozone complex has no magnetic moment, $\mathbf{m}(\ddot{O}\mathring{O}\ddot{O})=0$, magnetization stripes are absent in those compounds.\cite{2}

To summarize, there are commonalities and differences between the $Ae$-doped lanthanum cuprates from the first family and the oxygenated compounds from the second. What they have in common are neutral  \textit{oxygen atoms} in the $CuO_2$ planes. Where they differ is the \textit{site} (with corresponding relative charge) and magnetic moment of these lattice-defects---something not accounted for by the notion of ``doping the $CuO_2$ plane with holes.''

\section{CLOSING OF THE PSEUDOGAP IN DOPED $\mathbf{La_2CuO_4}$}
Pristine $La_2CuO_4$ is a Mott insulator of tetragonal crystal structure. Its $Cu^{2+}$ ions have a half-filled $3d_{x^2-y^2}$ orbital whose magnetic moment gives rise to antiferromagnetic (3D-AFM) order. 
When doped with $Ae$, the resulting holes in the $CuO_2$ planes settle pairwise at oxygen ions, endowing them with a magnetic moment, $\mathbf{m}(O^{2-}) \ne 0$, that compromises the 3D-AFM of the $Cu^{2+}$ ions. At $Ae$-doping $x = 0.02$ and $T = 0$, the 3D-AFM collapses and the so-called pseudogap phase ensues, with the appearance of charge-order and magnetization stripes, as mentioned. At a ``watershed'' value of doping, $\hat{x}$, the $CuO_2$ planes are saturated with doped holes (residing pairwise at defect $O$ atoms of the superlattice). This gives rise to a \textit{constant} incommensurability of stripes, $q_{CO, M}^{CuO_2}(x) = q_{CO, M}^{CuO_2}(\hat{x})$ for $x \ge \hat{x}$. Observed values of watershed doping are $\hat{x} = 0.125$ for $La_{2-x}Sr_xCuO_4$ 
and $\hat{x} = 0.16$ for co-doped $La_{2-x-z}Ln_zSr_xCuO_4,\;Ln = Nd, Eu; z = 0.4,\;0.2$. 
With further doping, $x > \hat{x}$, the resulting doped holes start residing 
pairwise at anion lattice-site $O$ atoms in the bracketing $LaO$ layers where they again form charge-order superlattices due to Coulomb repulsion. Their incommensurability is $q_{CO}^{LaO}(x) \propto \sqrt{x-\hat{x}}$. At the doping $x^*$ where the pseudogap closes,
the combined incommensurability of the bracketing 
$LaO$ layers equals that of the 
$CuO_2$ plane,\cite{3}
\begin{equation}
    2q_{CO}^{LaO}(x^*) = q_{CO}^{CuO_2}(x^*)\;,
\end{equation} 

\includegraphics[width=5.82in]{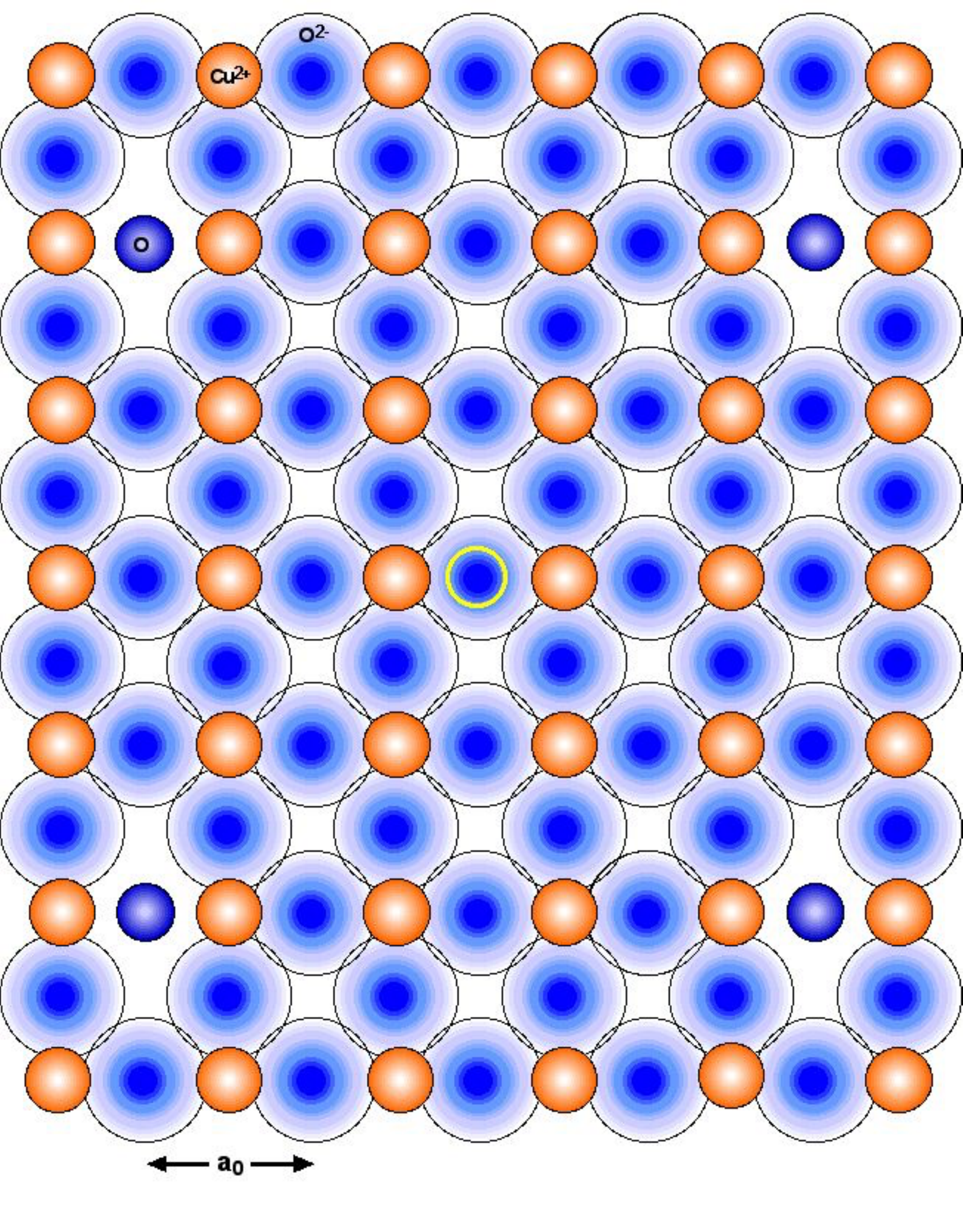}  \footnotesize 

\noindent FIG. 2. Positions of host-lattice ions and lattice-defect $O$ atoms in the $CuO_2$ plane, showing a unit cell of a \textit{commensurate} $O$ superlattice, $A_0 = 4 a_0$. It is roughly comparable to the incommensurate superlattice with $A_0 = 4.25 a_0$ when the pseudogap closes in $La_{2-x}Sr_xCuO_4$ or $A_0 = 3.75 a_0$ for  $La_{1.6-x}Nd_{0.4}Sr_xCuO_4$. Two interlaced superlattices of double spacing, $2A_0$, are formed by $O$ atoms in the bracketing $LaO$ layers --- one such $O$ atom is indicated by the yellow circle. The host lattice is shown in its \textit{unrelaxed} position. An extended and exclusive display of all three superlattices is shown in Fig. 3. 
\normalsize 

\noindent with $x^* = 0.18$ for $La_{2-x}Sr_xCuO_4$ and $x^* = 0.23$ for $La_{1.6-x}Nd_{0.4}Sr_xCuO_4$.

The positions of host-crystal ions and $O$ lattice defects are approximately shown in Fig. 2. For convenience of viewing, a close \textit{commensurable} superlattice of $q_{CO}^{CuO_2} = 0.25$ r.l.u.,  with superlattice spacing in the $CuO_2$ plane, $A_0 = 1/q_{CO}^{CuO_2} = 4a_0$, is displayed. The incommensurability of charge order in $La_{2-x}Sr_xCuO_4$ at the closing of the pseudogap is $q(x^*) = 0.235$ and that in $La_{1.6-x}Nd_{0.4}Sr_xCuO_4$ is $q(x^*) = 0.266$.\cite{3} The corresponding superlattice spacing is $A_0(x^*) = 4.25\;a_0$  and $A_0(x^*) = 3.75\;a_0$, respectively, deviating 

\includegraphics[width=5.9in]{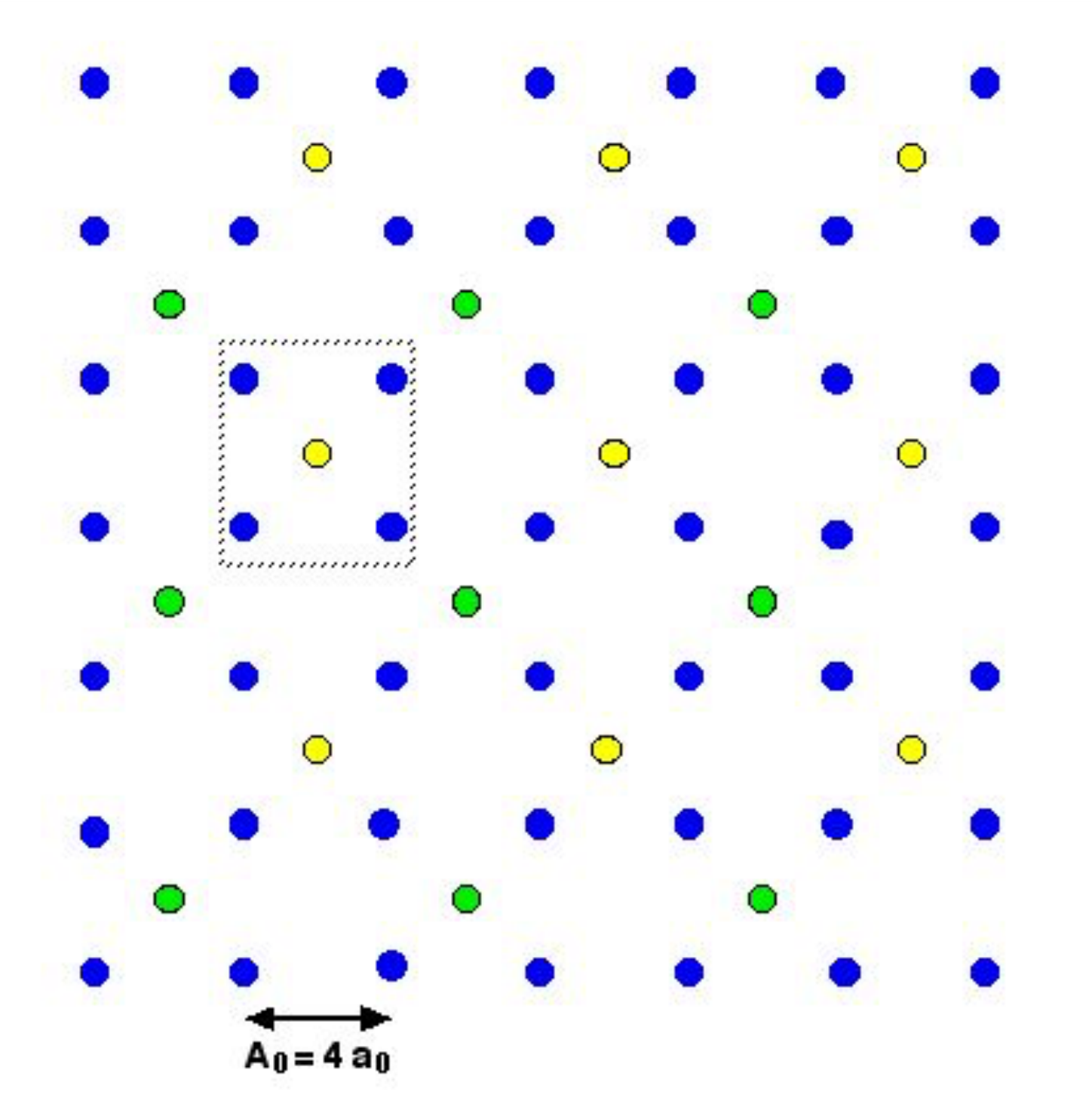}  \footnotesize

\noindent FIG. 3. Commensurable superlattices of $O$ atoms (size exaggerated) in the $CuO_2$ plane (blue), upper $\overline{LaO}$ layer (yellow), and lower $\underline{LaO}$ layer (green). Ions of the host lattice are not shown. The framed rectangle corresponds to Fig. 2.
The same symmetric pattern holds (slightly extended or contracted) for the incommensurable $O$ superlattices when the pseudogap closes in $La_{2-x}Sr_xCuO_4$ and $La_{1.6-x}Nd_{0.4}Sr_xCuO_4$.
\normalsize 

\noindent $\sim \pm 1/16 \simeq \pm 6 \%$ from the commensurate case shown in Fig. 2. 

Not distracted by the host lattice, Fig. 3 shows in an extended view the commensurate $O$ superlattice in the $CuO_2$ plane (blue dots) with spacing $A_0^{CuO_2} = 4 a_0$, and in the upper  $LaO$ layer (yellow  dots) and lower $LaO$ layer (green dots), each with spacing $A_0^{LaO} = 8 a_0$. More importantly, the \textit{same} symmetric pattern as in Fig. 3 holds when scaled to the spacing of incommensurable superlattices at the closing of the supergap, 
\begin{equation}
    A_0^{\overline{LaO}}(x^*)=A_0^{\underline{LaO}}(x^*)=2A_0^{CuO_2}(x^*)\;.
\end{equation}
Besides a small contribution from antiferromagnetism (AFM), the corresponding concentration of oxygen atoms in the $LaO $ layers and $CuO_2$ plane,
\begin{equation}
[O]^{\overline{LaO}^*}=[O]^{\underline{LaO}^*}=\frac{1}{4}[O]^{CuO_2^*}\;,
\end{equation}
gives rise to the relation between special doping values: $x^*-\hat{x} \simeq \frac{1}{2}\hat{x}$ or $x^* \simeq \frac{3}{2}\hat{x}$.\cite{3}

Charge order manifests as stripes in both cuprate families for low and medium hole doping. 
The unidirectional character of stripes in $La_{2-z-x}Ln_zAe_{x}CuO_4$  ($Ln = Nd, Eu;\; z =0,\; 0.4, \; 0.2$) is imposed by the low-temperature phases of the crystals. In these phases, $CuO_6$ octahedra are slightly tilted parallel to the planar crystal axes to reduce stress from lattice mismatch due to ion-size differences ($Ba^{2+} >Sr^{2+} \approx La^{3+}$), with the  same tilt for whole crystal domains. This creates a preference of charge-order and magnetization patterns in one direction over the orthogonal one, resulting in unidirectional stripes. The observed stripes-signal is maximal near $Ae$-doping $x=1/8=0.125$.
At larger doping levels the stripes-signal weakens till it disappears near ``optimal'' doping (for maximum $T_c$ of superconductivity), $p \approx 0.16$, as a cross-over of the charge order from (translational) stripes to nematicity---a $C_4$ rotational-symmetry-broken order---occurs. Qualitatively, the cross-over happens when closer proximity of lattice-defect $O$ atoms, due to increased doping, shifts the preference from unidirectional (tilt-related) alignment of far-separated (superlattice) $O$ neighbors to bidirectional fluctuations between closer-spaced $O$ neighbors. Whereas in stripes one direction of the square superlattice is perferred over the other, both directions of the superlattice manifest equally in the biaxial fluctuations of its nematic phase. 
With the high doping value at pseudogap closing,  $x^* = 0.18$ or $0.23$, the $O$ charge order clearly is in the nematic regime. Accordingly, its \textit{bidirectional} rather than unidirectional aspect is dominant.

\includegraphics[width=6in]{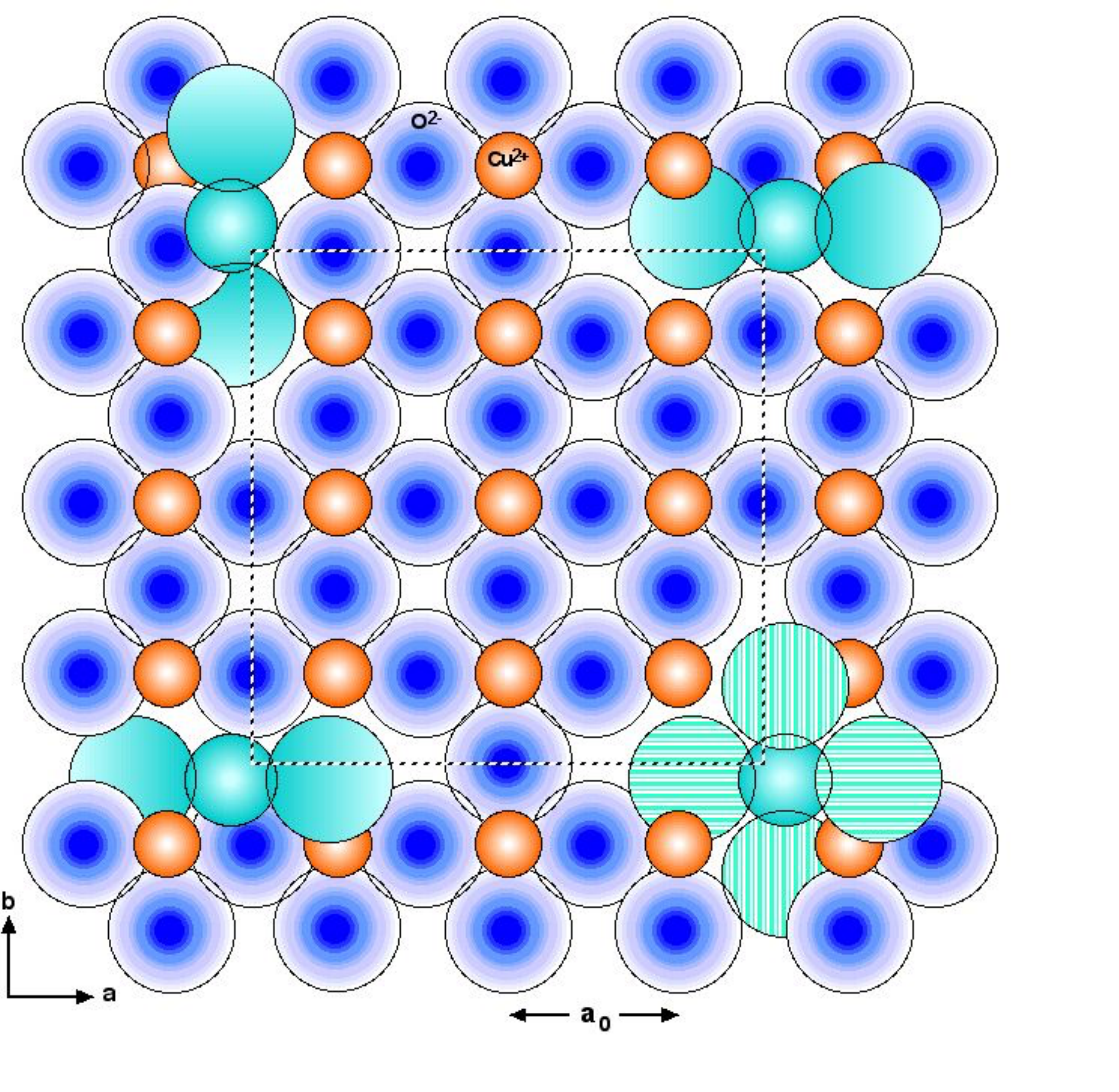}  \footnotesize 

\noindent FIG. 4. Atomic configuration of host-lattice ions in the $CuO_2$ plane (shown unrelaxed) of an oxygenated cuprate with a unit cell of the superlattice of ozone complexes  $\ddot{O}\mathring{O}\ddot{O}$ (turquoise color) at hole doping $p^*=0.19$ where the pseudogap closes (at $T=0$). With the charge order being in the nematic phase, all $\ddot{O}\mathring{O}\ddot{O}$ complexes fluctuate in both the $a$- and $b$-direction, as explicitly indicated near the lower right corner. The dashed $3a_0\times3a_0$ square facilitates recognition of the superlattice spacing, $A_0^* = 3.25a_0$.
\normalsize 

\section{CLOSING OF THE PSEUDOGAP IN OXYGENATED CUPRATES}
As no condition comparable to Eq. (1) is known for the closing of the pseudogap in oxygenated cuprates, a comparison of the atomic arrangement at the corresponding hole concentration $p^*$ (quantum critical point) in both cuprate families should be instructive.
From measurements of transport and spectroscopic properties,\cite{4,5,6,7,8,9} a hole concentration in the $CuO_2$ plane of $p^* = 0.19 \pm 0.01$ has been obtained for $YBa_2Cu_3O_{6+y}$, $Bi_2Sr_2CaCu_2O_{8+\delta}$ and $HgBa_2CuO_{4+\delta}$---a value that we want to pursue here.
However, two peculiarities of $Bi_2Sr_2CaCu_2O_{8+\delta}$ should be mentioned: (i) Its charge-order incommensurability is $q_{CO}^{CuO_2}(p) \simeq 0.31$ only up to $p \le 0.10$, followed by an abruptly drop to $q_{CO}^{CuO_2}(p) \simeq 0.27$, accompanied by X-ray scattering along a \textit{circle} of radius $\overline{q} = 0.27$ r.l.u.,
and slightly decreasing $q_{CO}^{CuO_2}(p)$ with further doping.\cite{10,11}
(ii) Besides $p^* \approx 0.19$, a value of $p^*=0.22$ has also been reported.\cite{12} 
Another peculiarity concerns the thallium cuprate  $Tl_2Ba_2CuO_{6+\delta}$ where a Lifshitz transition
has been observed at hole doping $p^{\between} \approx 0.25$, accompanied by charge-order of incommensurability $q_{CO}^{CuO_2}(p^{\between})= 0.31$.\cite{13}
(A Lifshitz transition is a change of the Fermi surface between hole-like and electron-like, centered in cuprates at the M = ($\frac{1}{2},\frac{1}{2}$) and $\Gamma = 0$ point of the Brillouin zone, respectively.)
Whereas in all other cuprates closing of the pseudogap coincides with a Lifshitz transition, $p^* = p^{\between}$, \textit{no} pseudogap phase is observed in $Tl_2Ba_2CuO_{6+\delta}$. These deviations from the previous trend are left aside until better understood.

 Since each interstitial oxygen atom, $\mathring{O}$, hosts two holes, the concentration of excess oxygen in the $CuO_2$ plane at the closing of the pseudogap is $[\mathring{O}]^* = p^*/2 = 0.095$. Assuming a square superlattice in the $CuO_2$ plane, $[\mathring{O}]^* = 1\mathring{O}/A_0^{*2}$, its spacing is $A_0^* = 3.24 a_0$ and its incommensurability $q_{CO}^{CuO_2}(p^*) = 0.31$ r.l.u. Note that these values agree with the spacing and, respectively, incommensurability of strings of ozone complexes, $\ddot{O}\mathring{O}\ddot{O}$, that give rise to unidirectional stripes in oxygenated cuprates (see Fig. 1). Whereas at low and medium oxygen doping/enrichment, the $\ddot{O}\mathring{O}\ddot{O}$ complexes aggregate to strings, forming unidirectional stripes, at higher oxygenation they start filling the $CuO_2$ plane bidirectionally to nematic patches, reaching \textit{completion} of a 2D $\ddot{O}\mathring{O}\ddot{O}$ superlattice with spacing $A_0^* \simeq 3.25 a_0$
 when, at hole doping $p^*$, the pseudogap closes.

\section{CONSEQUENCES OF THE SYMMETRIC CONFIGURATIONS}
Figure 5 shows the pseudogap temperature, $T^*$, of compounds from the two cuprate families as it linearly falls with low and medium hole doping. The common line for each family is a consequence of its common $O$ lattice defect. The different slopes of the lines result from the different $O$ defects between the families. 
Note that no kink is noticeable in the $T^*(x)$ line at the watershed values of the $Sr$-doped $La$-cuprates, $\hat{x} = 0.125$ and $\hat{x} = 0.16$, indicating an equal gradual influence from hole doping in the $CuO_2$ plane and the $LaO$ layers. On the other hand, the abrupt drop of both $T^*(x)$ lines at the respective $x^*$ values heralds a \textit{qualitative} change. What could that be?

In the case of the $Sr$-doped lanthanum cuprates, where lattice-site $O$ atoms spread, the qualitative change is imposed by the condition of Eq. (1) and manifests by the high symmetry of the interlaced superlattices in the $CuO_2$ plane and $LaO$ layers, Fig. 3. 
In the case of the oxygenated cuprates, where ozone complexes $\ddot{O}\mathring{O}\ddot{O}$ aggregate with a natural separation of $\sim 3.25 a_0$—--either as coaxial strings at low and medium hole doping or as nematic patches at higher $p$---the qualitative change occurs with the \textit{completion} of a 2D superlattice of ozone complexes, shown in Fig. 4 (by one of its unit cells). 
In both crystal 

\bigskip 

\includegraphics[width=6in]{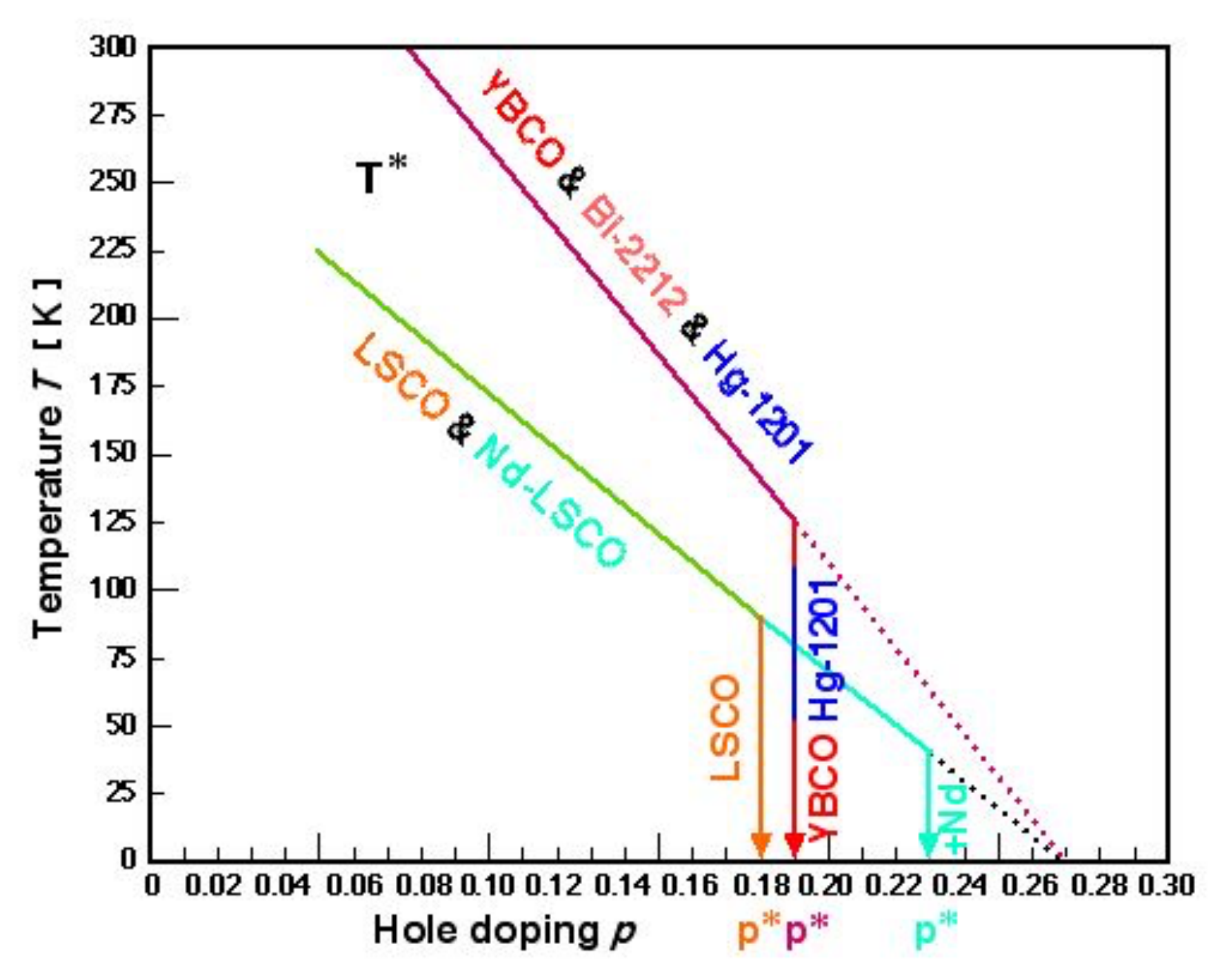}  \footnotesize 

\noindent FIG. 5. Pseudogap temperature $T^*$ of two doped lanthanum cuprates---$La_{2-x}Sr_xCuO_4$ (LSCO), $La_{1.6-x}Nd_{0.4}Sr_xCuO_4$ (Nd-LSCO) --- and three oxygenated cuprates --- $YBa_2Cu_3O_{6+y}$ (YBCO), 
$Bi_2Sr_2CaCu_2O_{8+\delta}$ (Bi-2212), $HgBa_2CuO_{4+\delta}$ (Hg-1201)---with abrupt drop when the pseudogap closes at $p^*$. The figure is reproduced and simplified from  figures (with data) in Refs. 6 and 9.
\normalsize 

\noindent  families each of these defect superlattices gives rise to a \textit{reciprocal} defect-superlattice in $q$-space, with corresponding \textit{defect-Brillouin zones}. Situated inside the first host-lattice Brillouin zone, $-\frac{1}{2} \le q_{a,b} \le +\frac{1}{2}$, the first defect-Brillouin zone extends about half in (linear) size, $-\frac{1}{4} \curlyeqprec Q_{a,b} \curlyeqprec +\frac{1}{4}$ for the $Sr$-doped lanthanum cuprates and roughly a third, $-\frac{1}{6} \curlyeqprec Q_{a,b} \curlyeqprec +\frac{1}{6}$, for the oxygenated cuprates.
The sudden occurrence of an additional (here, lattice-defect) Brillouin zone at doping $p^*$ can be expected to affect the bandstructure significantly. It should be considered in bandstructure calculations.

As known from basic solid-state physics, electron-electron scattering of quantum states near the Fermi surface involves \textit{umklapp} processes---backfolding of scattered quantum states by a reciprocal lattice vector $\mathbf{q_0}$. With the presence of an additional (defect) reciprocal lattice, \textit{competing} umklapp processes with the corresponding vector $\mathbf{Q_0}$ become feasible. They may give rise to the observed ``strange-metal'' phase with linear temperature dependence of electrical resistivity, $\rho \propto T$.


\begin{thebibliography}{16}
\bibitem{1} M. Bucher, ``Stripes in heterovalent-metal doped cuprates,'' arXiv:2002.12116v9

\bibitem{2} M. Bucher, ``Stripes in oxygen-enriched cuprates,'' arXiv:2010.06388v2

\bibitem{3} M. Bucher, ``Closing of the pseudogap in Sr-doped and Nd-codoped lanthanum cuprate,'' arXiv:2109.03046v2

\bibitem{4} J. L. Tallon and J. W. Loram,  Physica C \textbf{349}, 53 (2001).
\bibitem{5} S. Badoux, W. Tabis, F. Lalibert\'{e}, G. Grissonnanche, B. Vignolle, D. Vignolles, J. B\'{e}ard, D. A. Bonn, W. N. Hardy, R. Liang, N. Doiron-Leyraud, L. Taillefer, and C. Proust, Nature \textbf{531}, 210
(2016).
\bibitem{6} O. Cyr-Choini\`{e}re, R. Daou, F. Lalibert\'{e}, C. Collignon, S. Badoux, D. LeBoeuf, J. Chang, B. J. Ramshaw, D. A. Bonn, W. N. Hardy, R. Liang, J.-Q. Yan, J.-G. Cheng, J.-S. Zhou, J. B. Goodenough, S. Pyon, T. Takayama, H. Takagi, N. Doiron-Leyraud, and L. Taillefer, Phys. Rev. B \textbf{97}, 064502 (2018).

\bibitem{7} L. F. Lopes, J. P. Pe\~{n}a, M. A. Tumelero, J. Schaf, V. N. Vieira, and P. Pureur, Phys. Lett. A \textbf{383}, 2519 (2019).

\bibitem{8} Y. Li, V. Bal\'{e}dent, N. Bari\v{s}i\'{c}, Y. Cho, B. Fauqu\'{e}, Y. Sidis, G. Yu, X. Zhao, P. Bourges, and M. Greven, Nature \textbf{455}, 373 (2008).

\bibitem{9} N. Doiron-Leyraud, S. Lepault, O. Cyr-Choini\`{e}re, B. Vignolle, G. Grissonnanche, F. Lalibert\'{e}, J. Chang, N. Bari\v{s}i\'{c}, M. K. Chan, L. Ji, X. Zhao, Y. Li, M. Greven, C. Proust, and L. Taillefer, Phys. Rev. X \textbf{3}, 021019 (2013).

\bibitem{10} E. H. da Silva Neto, P. Aynajian, A. Frano, R. Comin, E. Schierle, E. Weschke, A. Gyenis, J. Wen, J. Schneeloch, Z. Xu, S. Ono, G. Gu, M. Le Tacon, and A. Yazdani, Science, \textbf{343}, 393 (2014).

\bibitem{11} F. Boschini, M. Minola, R. Sutarto, E. Schierle, M. Bluschke,6, S. Das, Y. Yang, M. Michiardi, Y. C. Shao, X. Feng, S. Ono, R. D. Zhong, J. A. Schneeloch, G. D. Gu, E. Weschke, F. He, Y. D. Chuang, B. Keimer, A. Damascelli, A. Frano, and E. H. da Silva Neto, Nat. Commun. \textbf{12}, 597 (2021). 

\bibitem{12} S. Benhabib, A. Sacuto, M. Civelli, I. Paul, M. Cazayous, Y. Gallais, M.-A. M\'{e}asson, R. D. Zhong, J. Schneeloch, G. D. Gu, D. Colson, and A. Forget, Phys. Rev. Lett. \textbf{114}, 147001 (2015).

\bibitem{13} C. C. Tam, M. Zhu, J. Ayres, K. Kummer, F. Yakhou-Harris, J. R. Cooper, A. Carrington, and S. M. Hayden, ``Charge density waves and Fermi-surface reconstruction in the clean overdoped cuprate superconductor
$Tl_2Ba_2CuO_{6+\delta}$,'' arXiv:2109.04279

\end{thebibliography}
\end{document}